\documentclass[aps,showpacs,prl,superscriptaddress,twocolumn]{revtex4-1}
\usepackage{booktabs}
\usepackage{graphicx,amssymb}
\usepackage{mathtools}
\usepackage{physics}
\usepackage{braket}
\usepackage{color}
\AtBeginDocument{
\heavyrulewidth=.08em
\lightrulewidth=.05em
\cmidrulewidth=.03em
\belowrulesep=.65ex
\belowbottomsep=0pt
\aboverulesep=.4ex
\abovetopsep=0pt
\cmidrulesep=\doublerulesep
\cmidrulekern=.5em
\defaultaddspace=.5em
}

\begin{document}

\title{Dual-frequency spin resonance spectroscopy of diamond nitrogen-vacancy centers in zero magnetic field}

\author{A. K. Dmitriev}
\thanks{These two authors contributed equally}
\affiliation{Ioffe Institute, St. Petersburg, Russia 194021}
\author{H. Y. Chen}
\thanks{These two authors contributed equally}
\author{G. D. Fuchs}
\affiliation{Cornell University, Ithaca, NY, USA 14853}
\author{A. K. Vershovskii}
\email{antver@mail.ioffe.ru}
\affiliation{Ioffe Institute, St. Petersburg, Russia 194021}

\date{\today}

\begin{abstract}

The methods for controlling spin states of negatively charged nitrogen-vacancy (NV) centers using microwave (MW) or radiofrequency (RF) excitation fields for electron spin and nuclear spin transitions are effective in strong magnetic fields where a level anti-crossing (LAC) occurs. A LAC can also occur at zero field in the presence of transverse strain or electric fields in the diamond crystal, leading to mixing of the spin states. In this paper, we investigate zero-field LAC of NV centers using dual-frequency excitation spectroscopy. Under RF modulation of the spin states, we observe sideband transitions and Autler-Townes splitting in the optically detected magnetic resonance (ODMR) spectra. Numerical simulations show that the splitting originates from Landau-Zener transition between electron spin $\ket{\pm1}$ states, which potentially provides a new way of manipulating NV center spin states in zero or weak magnetic field.

\end{abstract}
\maketitle

Optically detected magnetic resonance (ODMR) in the negatively charged nitrogen-vacancy (NV) center in diamond has been thoroughly studied over the past two decades. Many types of optically detected resonances, both magnetically dependent and magnetically independent, have been investigated, and various methods of controlling both electron and nuclear spins have been developed. These investigations have resulted in the development of high-resolution quantum magnetometers~\cite{1,2,3}, as well as methods of frequency stabilization~\cite{4}. Resonance techniques for controlling NV center electron and nuclear spins are considered to be highly suitable for quantum information processing because the relaxation time of these spins is long compared to many solid-state quantum objects~\cite{5,6}. Typically, either combined microwave (MW) and radiofrequency (RF) excitation~\cite{7,8,9,10,11,12,13,14}, or electron spin-echo envelope modulation methods~\cite{15} are used to address a chosen spin state, allowing for various multi-quantum resonances, e.g., as thoroughly studied in Ref.~\cite{12}. Various schemes of hole-burning using multi-frequency MW and RF excitation are discussed in Ref.\cite{14,16}; methods of all-optical excitation ODMR at two frequencies separated by the ground-state zero-field splitting are proposed in Ref.~\cite{17}.

Multi-frequency methods are most effective in strong magnetic fields (0.05 or 0.1 T), where a level anti-crossing (LAC)~\cite{5,6,12,16,17.1,18} in either excited or ground states occurs. A LAC between NV center spin states can also occur in weak~\cite{18} or zero magnetic fields mediated by external magnetic, intrinsic strain or electric field transverse to the N-V axis. The possibilities of using the zero-field LAC for controlling electron and nuclear spin states, or for exciting narrow resonances for metrological applications are not yet sufficiently explored. Here we present the results of our investigation of the zero-field LAC of NV centers in bulk diamond using dual-frequency (MW+RF) spin resonance spectroscopy. In contrast to normal ODMR spectra, we observe a splitting of the electron spin resonance. The frequency separation of the splitting is identical to the applied RF frequency and independent of magnetic field. When the RF field modulation frequency exceeds the linewidth of spin resonance, sideband transitions arise in the recorded spectra. We numerically simulate the dynamics using a Lindblad master equation and find excellent agreement with our experiment. From the simulation, we note that the splitting in the ODMR spectra originates from strong Landau-Zener transition between the spin $\ket{m_{s} =+1}$ and $\ket{m_{s} =-1}$ states mediated by RF modulation and the static intrinsic strain/electric field in diamond. This can also be described as Autler-Townes (A-T) splitting. Our approach constitutes a useful way of manipulating NV center spin states in the spin $\ket{m_{s}=\pm1}$ manifold at zero/low magnetic field.

The spin level structure of $^{3}A_{2}$ ground state of NV centers in external magnetic field is defined by the Hamiltonian~\cite{19,20,20.1}:

 \begin{equation}
\label{eq:Hamfull}
\begin{split}
H=&D(S^{2}_{z}-\frac{1}{3}\overrightarrow{S}^{2})-E_{x}(S^{2}_{x}-S^{2}_{y})+E_{y}(S_{x}S_{y}+S_{y}S_{x})
\\&+g_{s}\mu_{B}\overrightarrow{B}\cdot\overrightarrow{S}+A_{\parallel}S_{z}I_{z}+A_{\perp}(S_{x}I_{x}+S_{y}I_{y})
\\&+PI^{2}_{z}-g_{I}\mu_{N}\overrightarrow{B}\cdot\overrightarrow{I}, 
\end{split}
\end{equation}
where $\mu_{B}$= $h\cdot13.996\times10^{9}$~Hz/T is the Bohr magneton, $\overrightarrow{I}$is the $\mathrm{^{14}N}$ nuclear spin ($I$ = 1), $\overrightarrow{S}$ is the electron spin of NV center ($S=1$), $\mu_{N}= h\cdot7.622\times10^{6}$~Hz/T is the nuclear magneton, $D=(2\pi)2.87$~GHz and $E=\sqrt{E^{2}_{x}+E^{2}_{y}}$ is transverse zero-field splitting (ZFS) parameters. Depending on the local crystal strain and electric field, $E$ ranges from 0 for an unstressed diamond to up to more than $(2\pi)$10~MHz in highly strained diamond. $g_{s}=2.003$ and $g_{I}=0.403$ are electron and nuclear g-factors, $A_{\parallel}= -(2\pi)2.16$~MHz and $A_{\perp}= -(2\pi)2.7$~MHz are axial and transverse hyperfine splitting parameters, $P= -(2\pi)4.95$~MHz is the quadrupole splitting parameter. We denote eigenstates of the ground state $\ket{m_{s},m_{I}}$; for NV centers with the nitrogen isotope $\mathrm{^{14}N}$, both electronic and nuclear spin projections take values $m_{s}, m_{I}=0, \pm1$.

\begin{figure}
\includegraphics{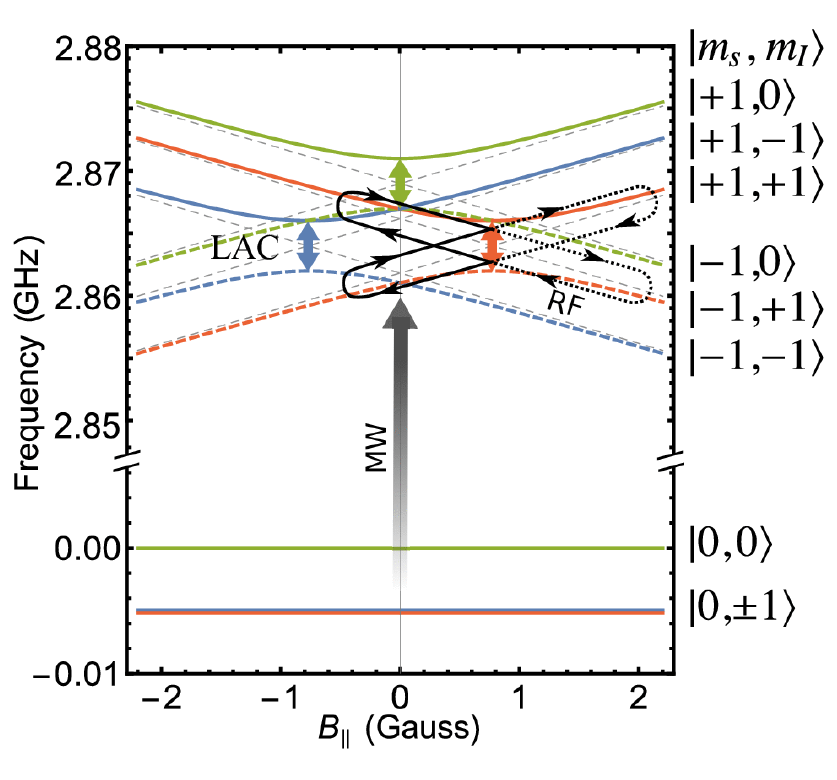}
\caption{(Color online) NV center ground-state splitting frequencies dependence on axial local magnetic field, calculated for a diamond crystal with transverse ZFS parameter $E$ = $(2\pi)$2~MHz. Thin dashed lines represent pure quantum states where $E=0$. Level anti-crossings mediated by transverse strain/electric field are marked by colored arrows. MW induced spin transitions preserve nuclear spin. The axial component of RF fields can lead to modulation of local magnetic field, $B_{\parallel}$, and hence a modulation of the spin state energies.}
\label{fig:1}
\end{figure}

The energy structure of NV centers in zero and weak magnetic fields is more complex than in strong ones (Fig.~\ref{fig:1}); it contains both level crossings and anti-crossings that are partially masked by the inhomogeneity of the crystal\textsc{\char13}s internal fields. Each NV center in a bulk sample is affected by local magnetic and electric fields. At $B\approx0$, the transverse component of the electric field combined with strain causes $\ket{m_{s}=\pm1}$ states to mix into superposition states, and transverse magnetic fields cause a second-order LAC~\cite{18}. The dependence of the magnetic field $B$ on the energy levels and corresponding frequencies in the low field regime is substantially nonlinear. This structure was investigated in detail in~\cite{18}.

\begin{figure}
\includegraphics{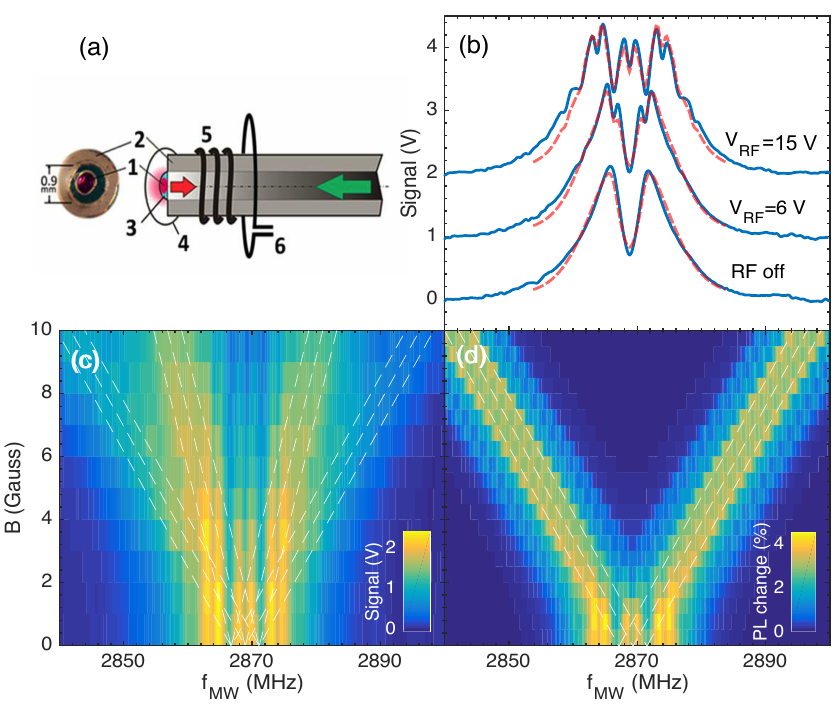}
\caption{(Color online) (a) View of the fiber end, and a schematic diagram of magnetometer sensor: 1$-$diamond crystal, 2$-$optical fiber, 3$-$transparent glue, 4$-$reflective coating, 5$-$MW antenna, 6$-$RF antenna. (b) ODMR spectra of NV centers taken with gradually increasing RF field amplitude at $f_{RF}$=5~MHz. Red dashed lines are a fit from numerical simulation. (c) ODMR spectra recorded at external field $B$ =(0 to 1)~mT along (1,1,1) crystal axis with additional RF excitation at $f_{RF}$=5 MHz. Spin transitions with a larger Zeeman splitting correspond to the axially aligned NV centers. (d) Numerical simulated results for NV centers oriented along (1,1,1) direction in (c). White dashed lines are guidelines for hyperfine spin transitions.}
\label{fig:2}
\end{figure}

The experiment was conducted at the Ioffe institute. The experimental setup was described in~\cite{21}: a synthetic diamond of SDB1085 60/70 grade (manufactured by Element Six, processed at the Lebedev Physical Institute) with dimensions $0.1\times0.3\times0.3$ mm was subjected to electron irradiation ($5\times10^{18}$ cm$^{-2}$) and subsequent annealing in Ar at 800 $^{\circ}$C for 2 hours. The diamond crystal was fixed with an optically transparent glue to the end of an optical fiber with a core diameter of 0.9 mm; the wide fiber aperture ensured effective collection of the photoluminescence (PL) signal at the cost of losing $\sim90\%$ of pumping light. The pumping and detection efficiency were further increased by covering the outer surfaces of the diamond and the end of the optical fiber with a non-conductive reflective coating (Fig.~\ref{fig:2}a).

The pumping beam ($\sim15$ mW at a wavelength of 532~nm) was focused on the second end of the fiber, and the PL signal was collected from the same end. Given an overall pumping efficiency of less than $10\%$, the pumping power was far below the optimum~\cite{10}. We perform dual-frequency spectroscopy by exciting ODMR in $B$=(0 to 1) mT using a MW drive field $f_{MW}$ in combination with additional RF field $f_{RF}$. We use low-frequency amplitude modulation of the MW field and synchronous detection at the modulation frequency in order to subtract the fluorescence background. All the experiments are done at room temperature.

Without RF field, the normal ODMR spectra (Fig.~\ref{fig:2}b) appears with two broad peaks in zero magnetic field. The splitting between the two peaks can be attributed to transverse static strain or electric field, $E$, intrinsic to the diamond~\cite{23}. After fitting to a simulated ODMR spectrum, we find an inhomogeneity in the range $E=0$ to $(2\pi)15$~MHz for the NV center ensemble in our sample, with an approximate Gaussian probability distribution, $p(E)=\frac{1}{\sqrt{2\pi}\sigma}$Exp$(-E^{2}/2\sigma^{2})$, where $\sigma$=$(2\pi)5$~MHz. We then apply RF field with gradually increasing amplitude and a fixed frequency, $f_{RF}=5$~MHz. The recorded ODMR spectrum (Fig.~\ref{fig:2}b) splits in each broad peak at $f_{s+}$ and $f_{s-}$, and the separation of the splitting is equal to the applied RF field frequency~\cite{22}, and is symmetrical about the zero field splitting: $f_{s+} -f_{s-} =f_{RF}, (f_{s+} +f_{s-})/2=D$. Comparing to the normal ODMR spectrum, there are increases in signal amplitude, both in between and outside the splitting in the RF dressed ODMR spectrum. They are around 5 MHz apart with respect to the peaks of the normal ODMR spectrum, which are later associated with the $1^{\text{st}}$ and $2^{\text{nd}}$ order sideband spin resonance transitions.

To interrogate the origin of the observed features, we then carry out experiments at both zero and weak ($<$~1~mT) magnetic field. We first investigate possible magnetic field dependence of the observed splitting, we apply external magnetic field (along (111) direction of the diamond lattice, Fig.~\ref{fig:2}c), $B$= (0 to 1)~mT, and record ODMR spectra with the same RF driving field. As $B$ increases, the splitting gradually fades while the separation of the splitting remains equal to the applied RF field frequency, independent of $B$. In a sufficiently strong $B$ field, spin resonances of two groups (the group with the larger splitting corresponds to the aligned NV centers) of NV centers appear due to the Zeeman effect. Each electron spin transition contains three peaks, corresponding to the hyperfine hybridization by $\mathrm{^{14}N}$ nuclear spin. 

\begin{figure}
\includegraphics{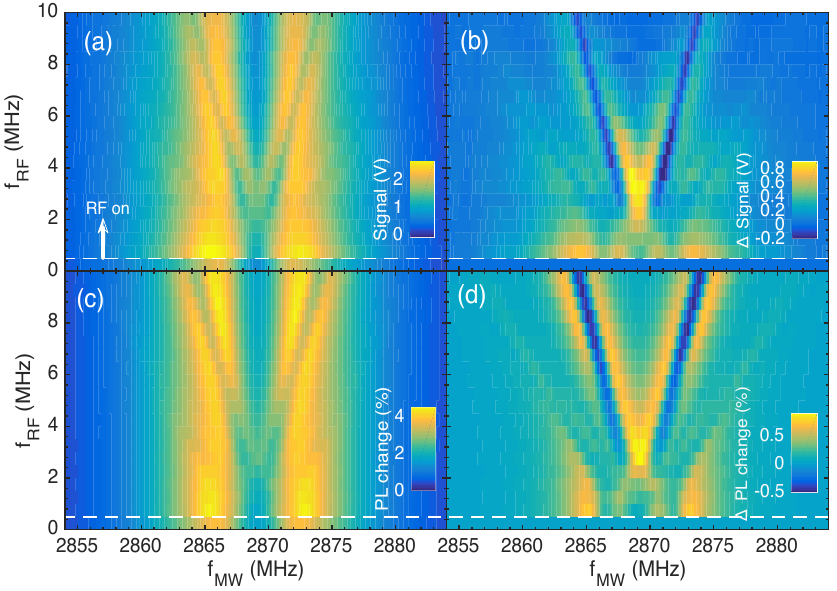}
\caption{(Color online)  (a) ODMR spectra recorded at zero external field at different radio frequencies $f_{RF}$ and a fixed amplitude of 6~V  (1~V corresponds to (2$\pi$)0.225~MHz in terms of Rabi frequency); the lowest row at zero frequency is the ODMR signal without RF driving. Splitting occurs in the ODMR spectra when RF field is applied, with a separation equal to the applied RF frequency. (b) Figure (a) subtracted from the normal ODMR spectrum when the RF is off. $1^{\text{st}}$ order and $2^{\text{nd}}$ order sideband spin transitions are present. (c) ODMR spectra taken with increasing RF field amplitude at a fixed frequency, $f_{RF}$=5~MHz. Sideband transitions up to $3^{\text{rd}}$ order are present. (c) and (d) are numerical simulations of the experiment, which agree well with the data.}
\label{fig:3}
\end{figure}

After excluding the effect of magnetic field, we focus on study in zero field. With fixed RF driving amplitude, we vary the applied RF field frequency from 0.5~MHz to 10~MHz. The resulting ODMR spectra are plotted in Figure \ref{fig:3}. To emphasize the effect of RF driving, in Fig.~\ref{fig:3}b, we subtract the spectra in Fig.~\ref{fig:3}a by the normal ODMR spectra without RF field. The differences in the RF dressed ODMR spectra reveal: 1) the splitting occurs for the entire RF frequency range (0.5 to 10~MHz) in a V shape, with a separation that grows linearly with the applied RF frequency; 2) sideband transitions are apparent outside the splitting as a result of spin level modulation of the RF field; 3) a LAC is observed in between $1^{\text{st}}$ order and $2^{\text{nd}}$ order sidebands. 

\begin{figure}
\includegraphics{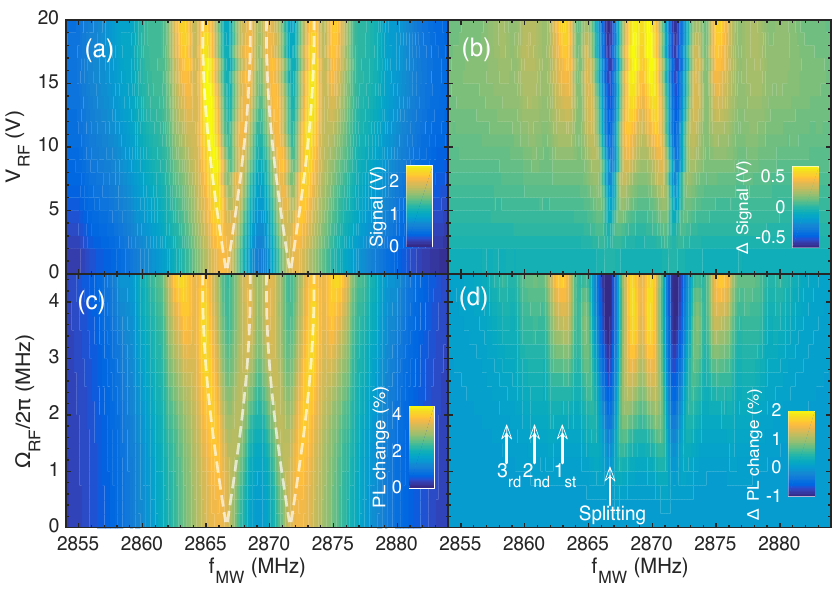}
\caption{(Color online)  (a) ODMR spectra taken with an increasing RF field amplitude at a fixed frequency, $f_{RF}$=5~MHz. (b): Figure (a) subtracted from the ODMR spectrum when RF is off. Sideband transitions up to $3^{\text{rd}}$ order are present. (c) and (d) are numerical simulations of the experiment. Overlay white dashed lines are theoretical prediction of the splitting from Eq.~(\ref{eq:Hamprime}), assuming $E$= (2$\pi$)2~MHz.}
\label{fig:amp}
\end{figure}

In the third experiment, we keep the RF frequency fixed at 5~MHz, and record ODMR spectra with increasing RF field amplitude (Fig.~\ref{fig:amp}). We observe the splitting grows first linearly and then sub-linearly as the applied RF field amplitude increases. Sideband transitions up to the $3^{\text{rd}}$ order are present in the probed range.

To interpret the experimental results, we ran numerical simulations using a 19$\times$19 Hamiltonian comprising 9 hyperfine levels in the ground states, 9 hyperfine levels in the excited states and a singlet level that is accessed via the intersystem crossing. We initialize the density matrix of the system in an unpolarized ground spin state, and evolve the system using a Lindblad master equation. Fluorescence of NV centers are calculated through the population in 9 excited hyperfine states that decay directly into the corresponding ground states, preserving electron/nuclear spin. Considering the inhomogeneity of the strain (or electric field) in our sample, we simulate dual-frequency ODMR spectra for NV centers of various strain, $E = 0$ to $(2\pi)15$~MHz, and average the spectra with a normal distribution of the strain $p(E)$. Using the following parameters: MW probe field Rabi frequency $(2\pi)$0.5 MHz, laser excitation optical Rabi frequency $(2\pi)$2~MHz, RF field Rabi frequency $(2\pi)$1.5~MHz, the simulated results are in excellent agreement with experimental data (Fig.~\ref{fig:3}c,d and Fig.~\ref{fig:amp}c,d).

\begin{figure}
\includegraphics{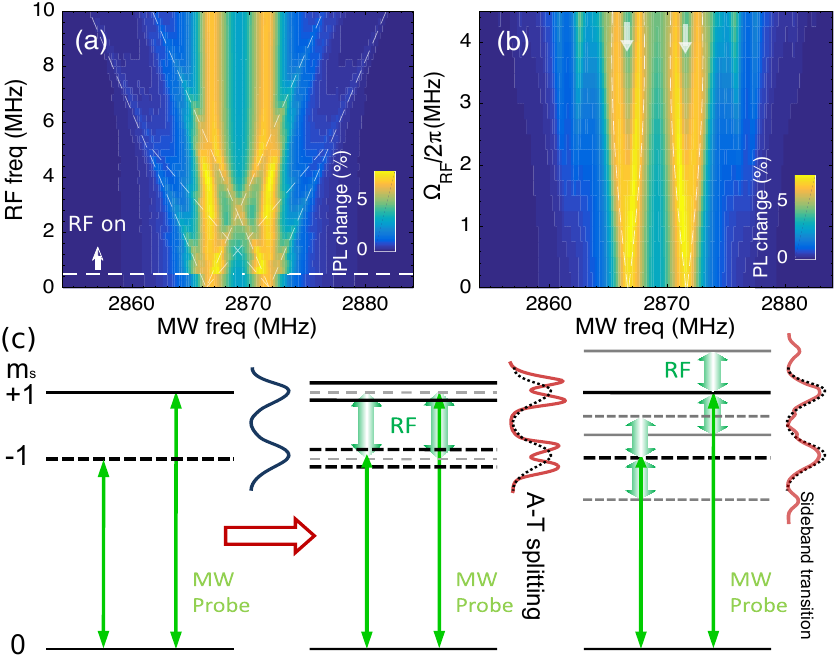}
\caption{(Color online) Numerically simulated dual-frequency ODMR spectra for a single NV center with transverse strain $E=(2\pi)$1.5~MHz. The RF driving field is kept at $\Omega_{RF}$=$(2\pi)$2~MHz in (a) and $f_{RF}$=5~MHz in (b). Sideband transitions and A-T splitting are labeled by white dashed lines and arrows. (c) Level diagrams of RF dressed spin states, showing effects of A-T splitting and sideband transitions.}
\label{fig:5}
\end{figure}

We now unravel the mechanism of splitting and sideband transitions in our experiment. For simplicity, we restrict the discussion to the $\{\ket{-1,+1}$, $\ket{+1,+1}\}$ basis for a single NV center. Figure \ref{fig:5} shows the numerically simulated dual-frequency ODMR spectra for a single NV center with transverse strain $E=(2\pi)$2~MHz. In the rotating frame of the MW field and in the absence of an external magnetic field, the Hamiltonian of the above states can be written as 
\begin{equation}
\label{eq:Ham}
\begin{split}
{H}&=
\begin{pmatrix}
-A_{\parallel}+\Omega_{RF}\text{cos}(2\pi f_{RF}t)  & E
\\ 
E & A_{\parallel}-\Omega_{RF}\text{cos}(2\pi f_{RF}t)
\end{pmatrix}
\end{split}
\end{equation}
where $\Omega_{RF}$ and $f_{RF}$ are amplitude and frequency of the RF driving field. The two levels are coupled by a transverse strain field, $E$. RF magnetic field modulates the energy of the two levels through the longitudinal Zeeman coupling. As the two levels approach each other, transitions can occur through Landau-Zener tunneling (LZT)~\cite{24}, as demonstrated in~\cite{25, 17.1} for a single NV center between $\ket{+1,0}$ and $\ket{0,0}$ states at finite field, with coupling provided by MW field. To explain the splitting, we apply the polaron transformation, such that the effective Hamiltonian becomes\cite{26, 26.1} 

\begin{equation}
\label{eq:Hamprime}
\begin{split}
{H}'&=
\begin{pmatrix}
-2\pi f_{RF}+\Delta  & \widetilde{\Omega}^{\dagger}\\ 
\widetilde{\Omega} & -\Delta 
\end{pmatrix},
\\
\widetilde{\Omega}&=-\Omega_{RF}\text{sin}2\theta\sum\limits_{n}  J_{n}(2\frac{\Omega_{RF}}{2\pi f_{RF}}\text{cos}2\theta)e^{in2\pi f_{RF}t}, 
\end{split}
\end{equation}
where tan2$\theta$=-$E/A_{\parallel}$,  $\Delta^{2}$=$A^{2}_{\parallel} +E^{2}$. $J_{n}$ is the Bessel function of the first kind. When RF field frequency meets the condition $2\pi f_{RF}=2\Delta$, a LZT spin transition occurs. When the RF diving amplitude is strong enough, the system is well-described by a dressed-states picture in which the RF field introduces a splitting $2\widetilde{\Omega}$ in the ODMR spectra (white dashed line trace in Fig.~\ref{fig:amp}). Taking into account the spin $\ket{m_{s}=0}$ state, we have a three-level system, with spin $\ket{m_{s}=\pm1}$ states coupled via transverse strain and RF field, probed by the MW transition from the spin $\ket{m_{s}=0}$ state. In the strongly-driven regime, this constitutes an Autler-Townes type structure. Due to RF field longitudinal level modulation of spin states, sideband transitions also occur in our spectra, which has been reported in \cite{12}. Applying the rotating frame and rotating wave approximation~\cite{27} (or Floquet theory~\cite{28}), $H$ becomes 
\begin{equation}
\label{eq:HamRWA}
\begin{split}
H_{RWA}&=
\begin{pmatrix}
-A_{\parallel}  & EJ_{n}(\frac{\Omega_{RF}}{2\pi f_{RF}})\\ 
EJ_{n}(\frac{\Omega_{RF}}{2\pi f_{RF}}) & A_{\parallel}-2n\pi f_{RF}
\end{pmatrix}.
\end{split}
\end{equation}
Sideband transitions of $n^{\text{th}}$ order arise at $nf_{RF}$ away from the unperturbed transition frequency.

We exclude other origins for the splitting seen in the experiment, for instance, transverse magnetic field from stray/earth field, RF driving of nuclear spin transitions, because transverse magnetic field only induces mixing between $\ket{\pm1,m_{I}}$ and $\ket{0,m_{I}}$ states, and nuclear spin flops driven by RF field occur only within states of the same electron spin. Neither is able to drive a transition between $\ket{-1,m_{I}}$ and $\ket{+1,m_{I}}$ states.

Note that transitions between $\ket{-1,m_{I}}$ and $\ket{+1,m_{I}}$ states of an NV center are magnetically forbidden, however, our results show that such transition can be driven through LZT using an RF field for a strained NV center in zero/weak magnetic field. With pulsed MW and RF sequence control, it is possible to coherently control the spin states for the $\ket{m_{s}=\pm1}$ levels with high fidelity. Given that strain/electric fields are prevalent in diamond crystals, RF control of spin $\ket{m_{s}=\pm1}$ states can be readily applied to a single NV center.

In conclusion, we perform dual-frequency spin resonance spectroscopy on NV centers using MW and RF field in zero/weak magnetic field, and report the detection of Autler-Townes splitting and sideband transitions in the recorded ODMR spectra. Numerical simulations reveals that the splitting is caused by strong Landau-Zener tunneling between $\ket{-1,m_{I}}$ and $\ket{+1,m_{I}}$ states, mediated by a zero-field LAC due to the transverse strain/electric field. This is a normally forbidden magnetic transition, thus demonstrating an alternative mechanism for manipulating NV center ground spin states.

\begin{acknowledgments}
Research support at Cornell University was provided by the Office of Naval Research (Grants No. N000141712290).

\end{acknowledgments}

\end{document}